\documentclass[prl,aps,reprint,twocolumn,superscriptaddress]{revtex4-2}

\usepackage{graphicx}
\usepackage{amsfonts,amssymb}
\usepackage{bbold}
\usepackage{color}
\usepackage{hyperref}
\usepackage{epstopdf, epsfig}
\usepackage{tikz}
\usepackage{mathtools}

\makeatletter
\newcommand{\thickhline}{%
  \noalign {\ifnum 0=`}\fi \hrule height 1pt
  \futurelet \reserved@a \@xhline
}
\makeatother
\newcommand{\bX}{{\ensuremath\boldsymbol{X}}}
\newcommand{\bu}{{\ensuremath\boldsymbol{u}}}

\newcommand{\cemefaddress}{Mines Paris, PSL University, CNRS, Cemef, Sophia Antipolis, France}
\newcommand{\inriaaddress}{Universit\'{e} C\^{o}te d'Azur, Inria, CNRS, Calisto team, Sophia Antipolis, France}
\newcommand{\inphyniaddress}{Universit\'{e} C\^{o}te d'Azur, CNRS, Institut de Physique de Nice, France}

\graphicspath{{./figs/}}

\begin{document}

\author{J\'er\'emie Bec} \affiliation{\inriaaddress}\affiliation{\cemefaddress}
\author{Christophe Brouzet} \affiliation{\inphyniaddress}
\author{Christophe Henry} \affiliation{\inriaaddress}

\title{Enhanced transport of flexible fibers by pole vaulting in turbulent wall-bounded flow}

\begin{abstract}
Long, flexible fibers transported by a turbulent channel flow sample non-linear variations of the fluid velocity along their length. As the fibers tumble and collide with the boundaries, they bounce off with an impulse that propels them toward the center of the flow, similar to pole vaulting. As a result, the fibers migrate away from the walls, leading to depleted regions near the boundaries and more concentrated regions in the bulk. These higher concentrations in the center of the channel result in a greater net flux of fibers than what was initially imposed by the fluid. This effect becomes more pronounced as fiber length increases, especially when it approaches the channel height.
\end{abstract}

\maketitle

The concentration and orientation of fibers in wall-bounded turbulent flows have been extensively studied in the past decades \cite{lundell2011fluid, voth2017anisotropic}. Earliest work has focused on numerical simulations of rigid fibers, either shorter~\cite{voth2017anisotropic,Oucheneetal2018,MichelArcen2021} or longer~\cite{Doquangetal2014} than the smallest length scale of the flow. These simulations have shown that the fibers tend to adopt a random orientation in the center of the channel, while aligning with the streamwise direction close to the walls. Near-wall coherent structures were found to play an important role, as fibers could accumulate in vortices, depending on their length and inertia, leading to large inhomogeneities in concentration and velocity distribution. Recent experiments~\cite{Dearingetal2013,Hakanssonetal2013,capone2017translational,Shaiketal2020,alipour2021long, alipour2022influence,BakerColetti2022,ShaikvanHout2023} confirmed these findings. Accumulation in near-wall structures can result from various effects, such as interactions with high-speed streaks~\cite{alipour2022influence}, turbophoresis (that causes inertial particles to migrate toward the wall), as well as interactions with boundaries, especially for rod-like objects~\cite{moses2001investigation, capone2017translational,Doquangetal2014,BakerColetti2022}. 
Still, these studies focused on rigid fibers much shorter than the channel height. It is expected that longer and more flexible fibers will be deformed by the flow, leading to new behaviors not yet fully explored.

Recent simulations have demonstrated that flexible fibers with inertia also tend to accumulate in the near-wall region~\cite{dotto2019orientation,dotto2020deformation}. The fibers considered there are longer than the smallest scale of the flow, but still much shorter than the channel height. To date, there are no studies on the behavior of flexible fibers with sizes comparable to the channel height and, to the best of our knowledge, they have only been investigated in homogeneous isotropic turbulence. In this context, laboratory experiments have characterized the rigid-flexible transition~\cite{brouzet2014flexible} and curvature statistics~\cite{Gayetal2018} of long flexible fibers. Long elastic chains have been simulated~\cite{picardo2018preferential, picardo2020dynamics}, suggesting that they tend to align with vortex filaments and remain trapped for extended periods. Other numerical studies on long flexible fibers in homogeneous isotropic turbulence have demonstrated their potential for measuring two-point statistics of turbulence~\cite{rosti2018flexible}, or exploring modulation mechanisms of turbulence~\cite{Olivierietal2022}. The aim of this study is to investigate the near-wall dynamics of elongated flexible fibers induced by collisions with boundaries, filling the gap in our understanding of their behavior in wall-bounded turbulent flows.

\begin{figure}[t!]
  \centering
  \includegraphics[width=0.5\textwidth]{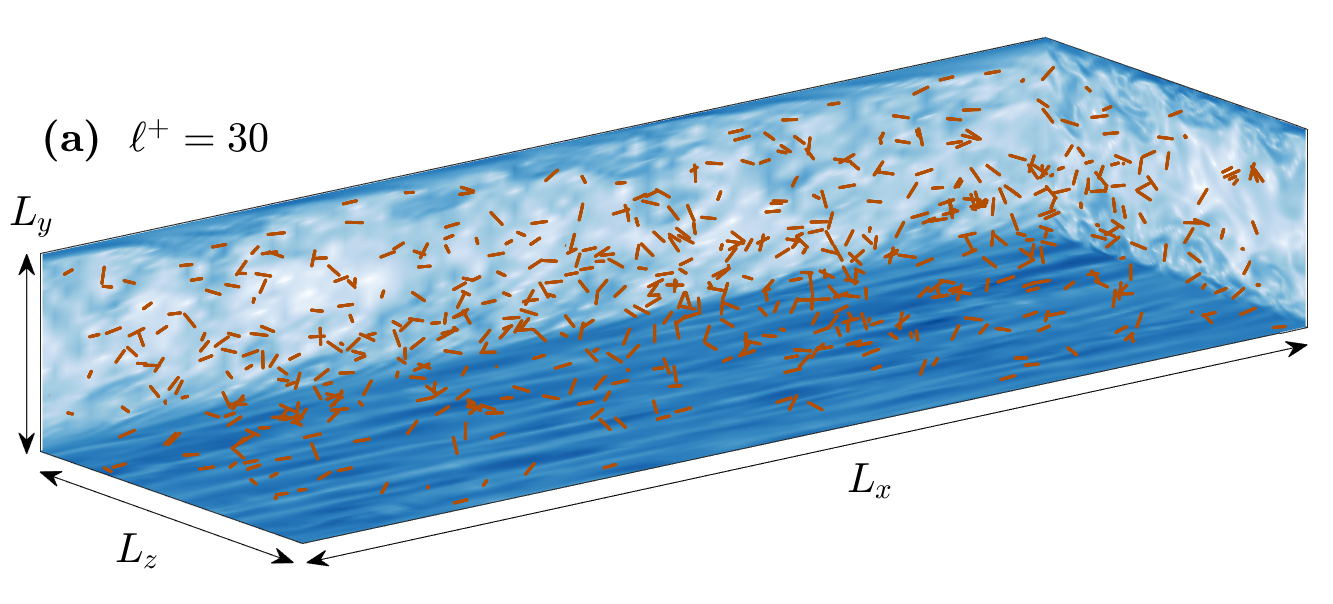}\\[-8pt]
  \includegraphics[width=0.5\textwidth]{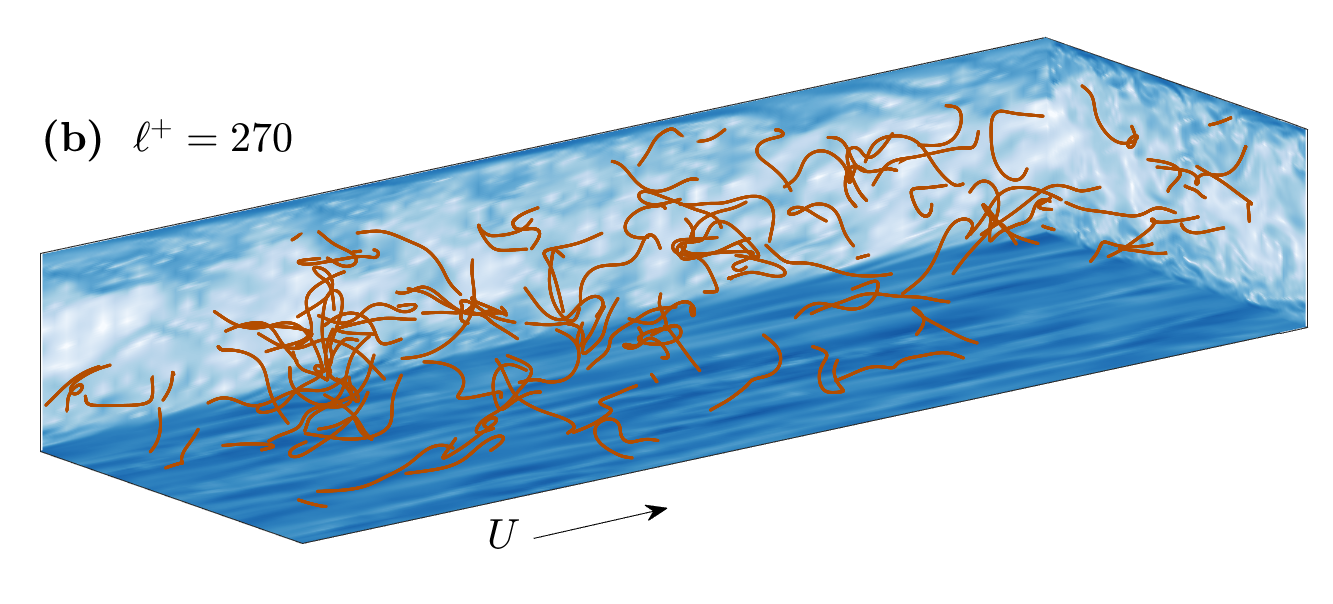}
  \vspace{-20pt}
  \caption{\label{fig:snap_fib_channel}\textit{Instantaneous distributions of fibers.} \textbf{(a)}~$\ell^+=30$ and \textbf{(b)}~$\ell^+=270$. The colored panels show the modulus of fluid vorticity in the planes $x=L_x$, $y=-L_y/2=h$ and $z=0$. Note that near-wall structures or streaks are visible close to the plane~$y=-h$. The mean flow is from left to right.}
\end{figure}

In this study, our focus is on long, inextensible, and inertialess fibers immersed in a channel flow with moderate Reynolds number. We consider a one-way coupling between the fibers and the flow~\cite{Elgobashi2006}, which means that the fibers do not produce any feedback on the flow nor interact with one another. The over-damped slender-body theory is used to describe the fibers, whereby each fiber, with length $\ell$ and thickness $a$, is parametrized by its arc-length $s\in[-\tfrac{\ell}{2},\tfrac{\ell}{2}]$, and its position $\bX(s,t)$ is given by:
\begin{eqnarray}
\partial_t\bX &=& \bu(\bX,t) + \frac{c}{8\pi\rho_\mathrm{f}\nu} \mathbb{D}\left[\partial_s(T\partial_s
  \bX) - E\partial_s^4 \bX \right], \nonumber \\ 
  \mbox{with}&& \mathbb{D} = \mathbb{1} +
   \partial_s\bX \partial_s\bX^{\mathsf{T}} \quad\mbox{and}\quad |\partial_s \bX| = 1.
    \label{eq_SBT1}
\end{eqnarray}
The tension $T(s,t)$ is here a Lagrange multiplier associated to the inextensibility constraint, $E = YI$ is the fiber's bending modulus (with $Y$ the Young modulus and $I=\pi a^4/64$ the fiber's moment of inertia), $\rho_{\rm f}$ the fluid density, $\nu$ its kinematic viscosity and $c= -[1+2\,\text{log}(a/\ell)]$ a parameter related to the shape~\cite{allende2018stretching}. Equation~\eqref{eq_SBT1} involves the fluid velocity~$\bu(\bX,t)$ along the fiber. It is obtained in our numerical simulations from the open-source spectral code \textit{Channelflow 2.0}~\cite{channelflow}, used to solve the incompressible Navier--Stokes equation in a rectangular channel of size $L_x=4\pi h$, $L_y=2 h$, $L_z = 4\pi h/3$ (see Fig.~\ref{fig:snap_fib_channel}), periodic in the $x$ and $z$ directions and with no-slip boundary conditions at $y=\pm h$. The fluid flow is integrated with a resolution of $128\times129\times128$ collocation points. It is maintained in motion by imposing a constant bulk velocity, such that the friction Reynolds number is equal to $\textrm{Re}_\tau = u_\tau\,h/\nu \approx 180$ (with $u_\tau$ the friction velocity). From now on, we will be using classical wall units (denoted by a superscript~$+$), where time and length scales are non-dimensionalized by $\tau_\nu = \nu/u_\tau^2$ and $\delta_\nu = \nu/u_\tau$.

The dynamics of fibers in Eq.~\eqref{eq_SBT1} is integrated numerically by a second-order finite-difference scheme~\cite{tornberg2004simulating, allende2018stretching}. To prevent them from crossing the walls at $y=\pm h$, we adopt a soft-boundary approach where the fluid velocity outside the domain is replaced by $\bu = (0,v,0)$, with $v=-\gamma\,(y+h)$ for $y<-h$ and $v = -\gamma\,(y-h)$ for $y>h$. The parameter $\gamma$ is adjusted to the time step $\Delta t$ as $\gamma = 0.5/\Delta t$. In our simulations, fiber lengths take values in the range $\ell^+=\ell/\delta_\nu\in[15,270]$, \textit{i.e.} $\ell/h \in[0.08,1.5]$. Each fiber is discretized with a given number of grid points $N_s$ along its arc-length, chosen to keep a constant step size $\delta s^+=\ell^+/N_s  \approx0.9$. The number of simulated fibers $N_{\rm fib} \in [100,1250]$ decreases with increasing length $\ell^+$ for the sake of computational time.

In the model we are using, inextensible and inertialess fibers are characterized by two length scales. The first scale is their extension length $\ell$, which is expressed in wall units as $\ell^+ = \ell/\delta_\nu$. This value indicates how far into the turbulent boundary layer the fibers can extend. The second scale is the elastic length $\ell_{\rm E} = \left[c\,E \tau_\nu / (8\pi\,\rho_\mathrm{f}\,\nu) \right]^{1/4}$, which is maintained constant to $\ell_{\rm E}^+ \approx 7.1$ in our simulations. It is determined by balancing in Eq.~\eqref{eq_SBT1} the bending rate of the fibers, given by~$cE/(8\pi\rho_\mathrm{f}\nu\ell^4)$, with the typical shear rate $\tau_\nu^{-1}$ of the channel flow. Note that the definition of elastic length introduced in~\cite{brouzet2014flexible} for homogeneous isotropic turbulence is recovered when using the Kolmogorov time scale~$\tau_\eta$ instead of~$\tau_\nu$. The dimensionless flexibility of the fibers, defined as  $\mathcal{F} = \ell/\ell_{\rm E}$, measures their ability to bend near the walls of the channel, with smaller values of $\mathcal{F}$ indicating more rigid fibers.  In order to examine how flexible fibers behave in comparison to rigid rods (which are formally defined as having $\ell^+ = 0$ and $\mathcal{F}=0$), we also integrated the Jeffery equations for threadlike ellispoids in the same flow along tracer trajectories. Figure~\ref{fig:snap_fib_channel} shows two snapshots of the fiber distribution with different lengths. Upon qualitative inspection, it is apparent that short, rigid fibers remain very straight, whereas long, flexible fibers bend into complex shapes.

\begin{figure}[t!]
  \includegraphics[width=0.47\textwidth]{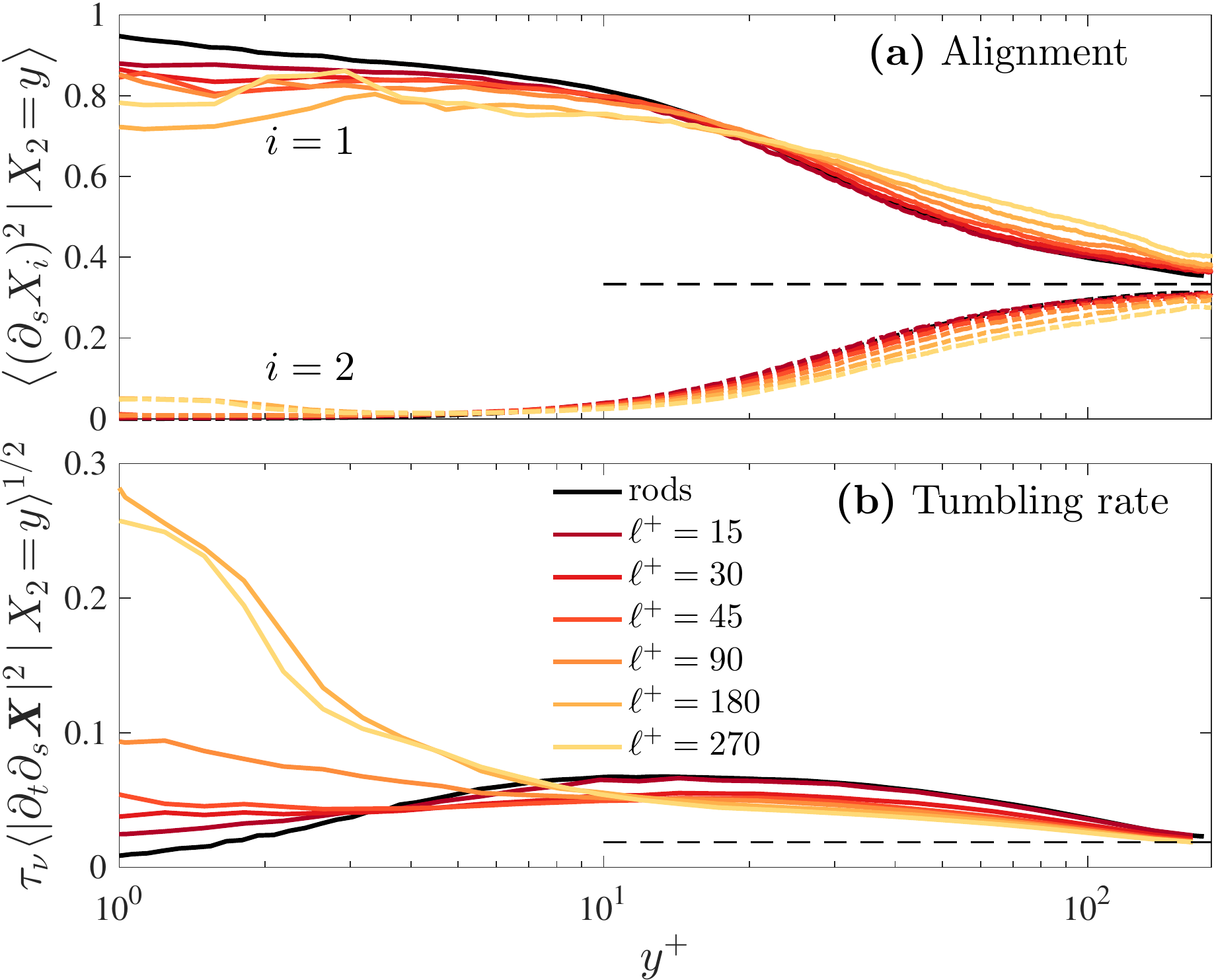}
  \vspace{-8pt}
  \caption{\label{fig:orientation_fibers}\textit{Near-wall alignment and tumbling.} \textbf{(a)}~Mean-square components of the   tangent $\partial_s\bX$ to the fiber in the streamwise (solid lines) and wall-normal (dashed lines) directions, as a function of the distance $y^+$ to the wall  and for various fibers lengths $\ell^+$. The horizontal dashed line corresponds to an isotropic orientation $\langle (\partial_s X_i)^2\rangle=1/3$.  \textbf{(b)}~Root-mean square amplitude of the tumbling rate $\partial_t\partial_s \bX$, again conditioned on $y^+$.  The dashed line shows here the value $\langle |\partial_t \partial_s \bX |^2\rangle\approx0.1/\tau^2_\eta$ expected for rods in homogeneous isotropic turbulence  ~\cite{Parsaetal2012,voth2017anisotropic}, where the Kolmogorov time $\tau_\eta=(\nu/\varepsilon)^{1/2}$ is computed using the bulk-flow turbulent dissipation rate~$\varepsilon$.}
\end{figure}

We first report results on the statistics of the fiber orientation and curvature as a function of their distance to the wall. Figure~\ref{fig:orientation_fibers}(a) represents the average orientation of the local tangent vector to the fiber conditioned on the distance $y^+$ to the nearest boundary. Only the streamwise ($x$) and wall-normal ($y$) components are represented, the spanwise component ($z$) being deduced from the inextensibility condition $|\partial_s\bX|^2=1$. 
Our findings confirm the previously-observed behavior of stiff rods ($\ell^+=0$, black lines) that tend to align with the mean flow near the boundaries and have a random orientation in the bulk~\cite{voth2017anisotropic}. Finite-size, flexible fibers show a qualitatively similar trend. However, their preferential orientation with the mean flow persists at larger distances, in particular for the longer fibers. Notably, longer fibers show a non-zero alignment with the wall-normal direction in the viscous sublayer, as observed for flexible fibers with inertia~\cite{dotto2019orientation}.

Regarding angular velocities, there are several ways to define a tumbling rate for flexible particles. We measure here the amplitude $|\partial_t\partial_s \bX|$ of the time derivative of the tangent vector along the fiber arc-length. This definition retrieves when $\ell\to0$ the well-known tumbling rate of rigid fibers $|\mathrm{d}\boldsymbol{p}/\mathrm{d}t|$, where $\boldsymbol{p}$ is the rod's orientation~\cite{voth2017anisotropic, pujara2021shape}. Figure~\ref{fig:orientation_fibers}(b) displays the root-mean-square value of the fibers tumbling rate conditioned on the wall-normal distance~$y^+$. In the channel center, they are all comparable and approach the value expected for rods in homogeneous isotropic turbulence~\cite{Parsaetal2012,voth2017anisotropic}. Interestingly, we observe that the smallest fibers with $\ell^+=15$ tumble similarly to rigid rods, except in close proximity to the wall, at distances $y^+\lesssim \ell^+/2$, where tumbling is slightly enhanced compared to rods. For intermediate sizes, tumbling is depleted far from the boundary but enhanced close to the wall. This effect becomes significantly more pronounced for the longest fibers ($\ell^+=180, 270$), where tumbling is drastically amplified in close proximity to the wall. These strong angular velocities may result from wall  contacts~\cite{BakerColetti2022} and, as we will see, give rise to important motions away from the boundary.

\begin{figure}[h!]
	\centering
 	\includegraphics[width=0.46\textwidth]{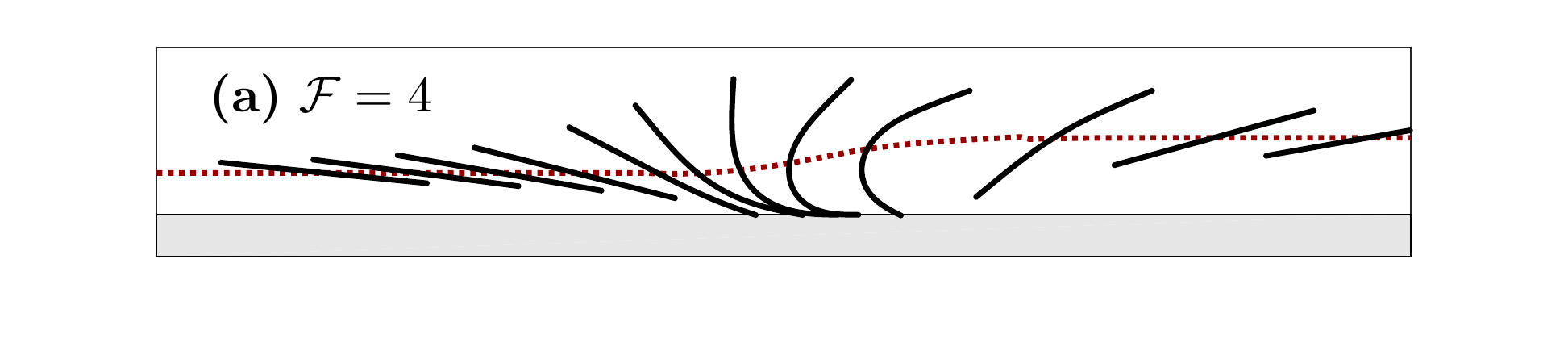}\\[-13pt]
  	 \includegraphics[width=0.46\textwidth]{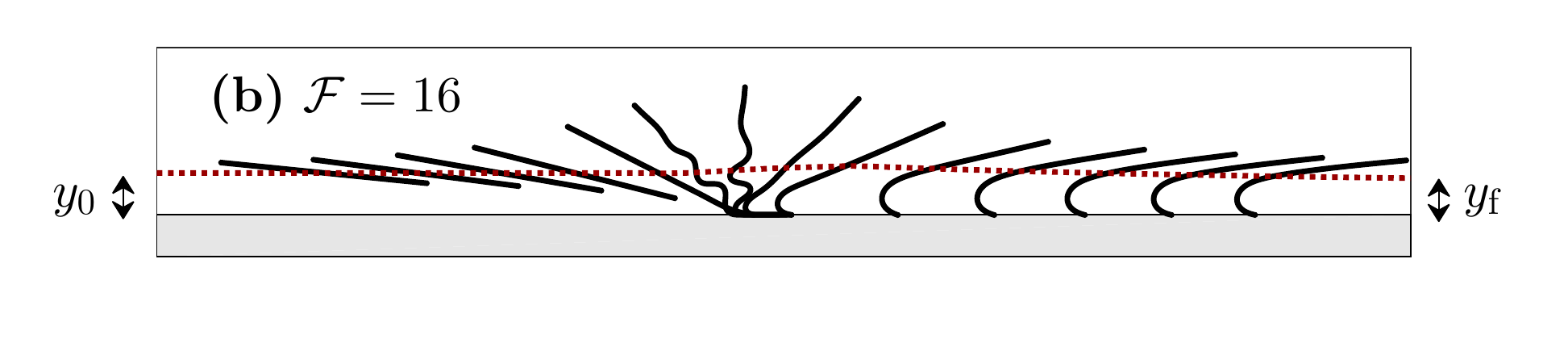}\\[-10pt]
	 \includegraphics[width=0.45\textwidth]{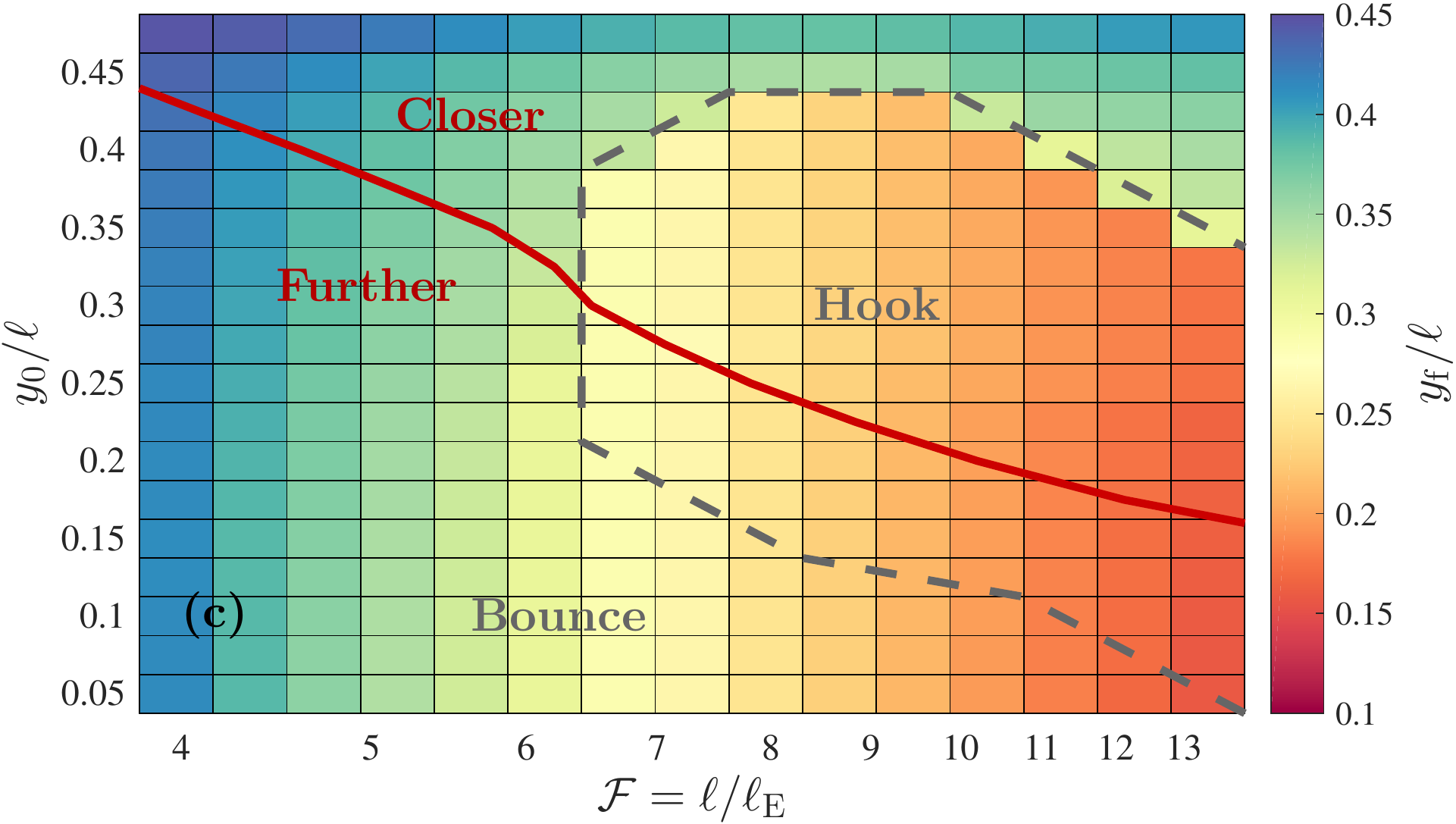}
  \vspace{-8pt}
  \caption{\label{fig:bouncing}\textit{Fibers tumbling in a pure shear flow $\bu = \sigma\,y\,\hat{\boldsymbol{x}}$}. \textbf{(a)}~Rigid fiber  bouncing on the wall. \textbf{(b)}~Flexible fiber ending up in a hook position. The red dotted lines are the center of mass trajectories. \textbf{(c)}~Phase diagram showing the fiber's final distance to the wall $y_{\rm f}$ as a function of the flexibility $\mathcal{F}$ and of its initial distance $y_0$. The dashed line separates bouncing and hook configurations. The solid line separates trajectories ending up further to the wall from those approaching it.}
\end{figure}

Insights into the near-boundary dynamics can be gained by considering a simple two-dimensional model where the fiber is confined in the $(x,y)$ plane and the flow over the wall is a pure linear shear flow $\bu = \sigma\,y\,\hat{\boldsymbol{x}}$. In this scenario the elastic length directly reads $\ell_{\rm E} = \left[c\,E / (8\pi\,\rho_\mathrm{f}\,\nu \sigma) \right]^{1/4}$. If a fiber is initially straight but slightly slanted downward, it will undergo a tumbling motion following a Jeffery orbit. If the fiber's center of mass is initially within a distance $\ell/2$ from the wall, the fiber will touch the boundary during this rotating motion. Upon collision, it will fold and the resulting displacement will be strongly influenced by its elasticity. Figure~\ref{fig:bouncing}(a) illustrates that when the fiber is rather rigid it bends and is propelled further from the wall. This effect, comparable to pole vaulting, is responsible for the fiber finishing its course at a higher height than initially. Similar pole-vaulting effects have been observed experimentally for rod-like particles \cite{moses2001investigation, capone2017translational,alipour2022influence,BakerColetti2022}. On the other hand, when the fiber is more flexible, as shown in Fig.~\ref{fig:bouncing}(b), it buckles upon contact with the wall and remains folded, ending its course in a hook shape. As a result of this folding, the final height of the fiber's center of mass can be lower than before tumbling occurred.

\begin{figure}[h!]
	\centering
 	\includegraphics[width=0.5\textwidth]{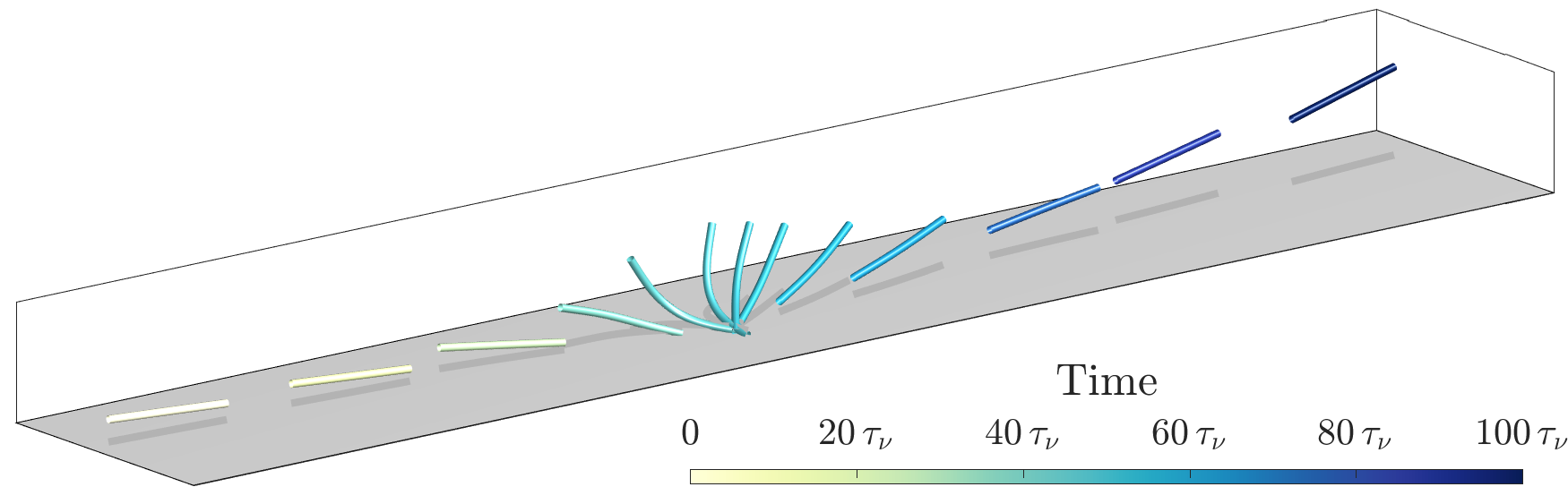}
  \vspace{-8pt}
  \caption{\label{fig:turb_jump}\textit{Near-wall tumbling of a fiber of length $\ell = 0.25\,h \approx 45\,\delta_\nu$ in the turbulent channel flow, leading to an instance of pole-vaulting.} Different colors correspond to various instant of time, expressed in viscous units.}
\end{figure}

The outcome of a near-wall tumbling event depends upon two parameters: the local non-dimensional flexibility $\mathcal{F}$ and the initial distance $y_0$ of the fiber from the wall. The occurrence of a pole-vault effect is intricately determined by these parameters, as demonstrated in the phase diagram of Fig.~\ref{fig:bouncing}(c). After a tumbling motion near the wall, fibers can end up either closer to or further away from the boundary. This depends whether the parameters are above or below the red solid line that  behaves approximately as $y_0\propto - \ell \, \log \mathcal{F}$. The gray dashed line is the separatrix between the two types of near-wall motion: bounce or hook. It confirms that the hook shape only occurs for fibers that are flexible enough. Yet, both types of motion can bring fibers closer or further to the wall. In actual turbulent boundary layers, the situation is even more complex due to local fluctuations of the shear and to the presence of near-wall structures. Furthermore, we consider fibers that are longer than the viscous scale~$\delta_\nu$, so their tumbling motion involves scales falling in the turbulent logarithmic layer. Nevertheless, pole-vaulting during near-wall tumbling seems a frequent occurrence in turbulent channel flow. Figure~\ref{fig:turb_jump} serves as an illustrative example, wherein the fiber is bent in three dimensions but still undergoes a strong enough impulse to migrate away from the wall. As we will now see, these interactions with the boundaries are the underlying cause of non-uniform fiber distributions in the channel, ultimately leading to the enhancement of their transport.

\begin{figure}[t]
  \centering
 \includegraphics[width=0.47\textwidth]{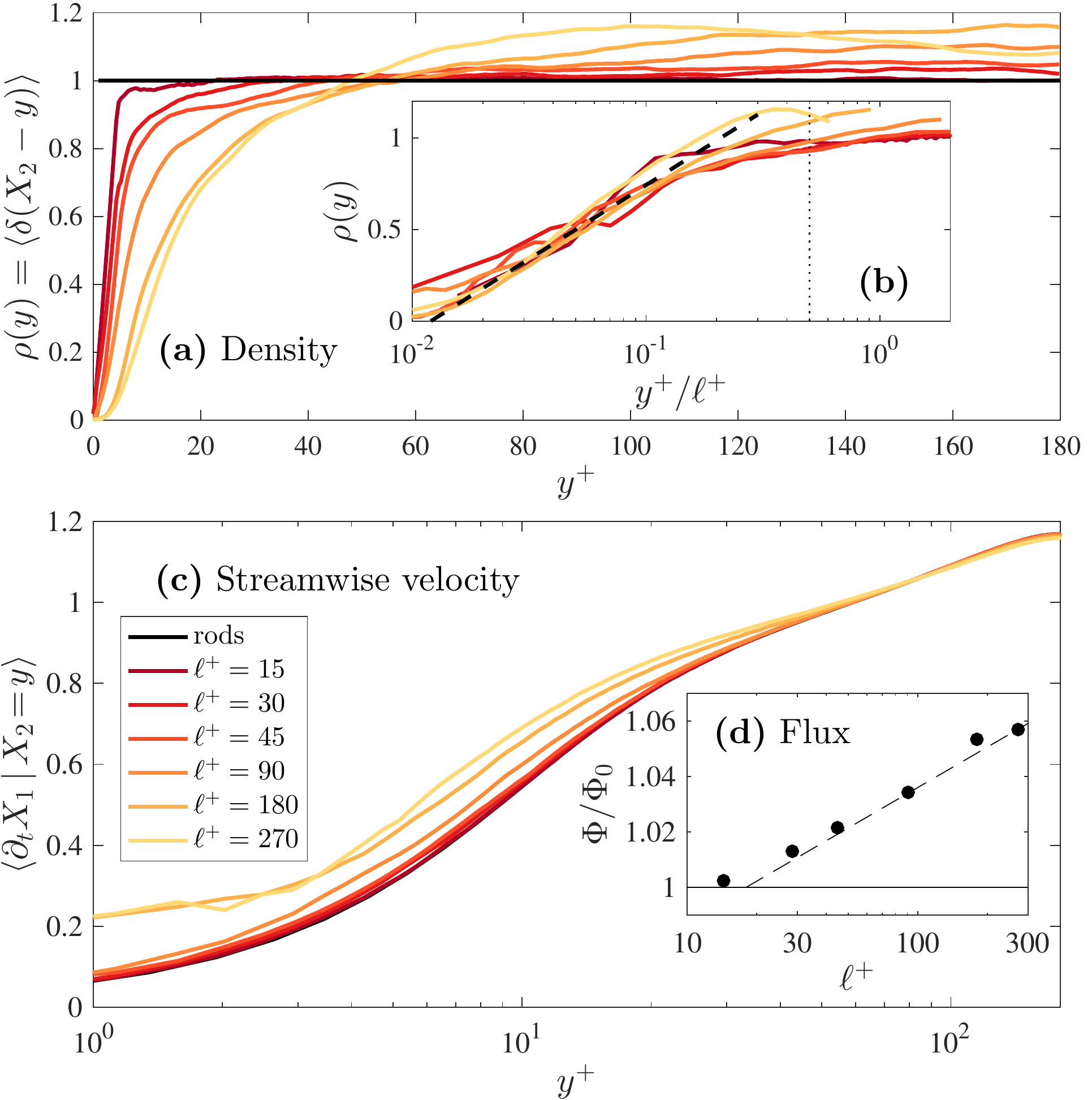}
  \caption{\label{fig:concentration_channel} \textit{Inhomogeneous fiber distributions and velocities.} \textbf{(a)}~Average density as a function of the distance to the wall for the various fiber lengths. \textbf{(b)}~Same, plotted as a function of $y^+/\ell^+$. The dashed line shows a behavior $\propto\log\ell^+$. \textbf{(c)}~Average fiber velocity in the streamwise direction, as a function of the wall-normal coordinate. \textbf{(d)}~Fiber fluxes normalised to that of the fluid. The dashed line represents a log behavior.}
\end{figure}

Building upon prior observations, we now present our findings on the spatial distribution and velocities of fibers.  Figure~\ref{fig:concentration_channel}(a) depicts the average concentration as a function of the wall-normal distance $y^+$. A depletion in the boundary layer is readily apparent, which becomes more pronounced with an increase in fiber length. As a result, fibers tend to accumulate far from the boundaries, towards the bulk of the channel flow. The inset~(b) of Fig.~\ref{fig:concentration_channel} shows the same data, but in semi-logarithmic coordinates and as a function of the ratio between the wall-normal distance and the fiber length $y^+/\ell^+$. Notably, the data seem to collapse onto a master curve, suggesting that fiber length is indeed the appropriate scale to characterize boundary depletion and corroborating the mechanisms described previously. 

A higher concentration of fibers in the bulk region, where the flow speed is higher, may suggest that fibers are transported more efficiently by the flow. However, as the fibers are not expected to follow exactly the fluid due to their important length, it is essential to examine first their velocity in the streamwise direction~$x$. Figure~\ref{fig:concentration_channel}(c) shows this mean velocity, with statistics conditioned on $X_2(s,t)=y$ and therefore not carrying any information on the concentration. One observes an increase in velocity close to the boundary and a minor decrease in the bulk, both of which become more pronounced with fiber length. The increase close to the walls could be attributed to the concentration of fibers in high-velocity streaks~\cite{Doquangetal2014,voth2017anisotropic,dotto2019orientation}. However, a long fiber in a channel flow can sample different regions of the flow simultaneously. Thus, segments of the fiber located at different distances from the wall experience different average fluid velocities. As the fiber is inextensible, the segments located in the channel center with a higher mean fluid velocity tend to pull the whole fiber, while those near the walls with a lower velocity slow it down. This mechanism results in a higher fiber streamwise velocity than the fluid near the walls and lower in the center, as observed in Fig.~\ref{fig:concentration_channel}(c).

By combining the concentration and velocity measurements to obtain the average flux in a $(y,z)$-plane, we find that these effects ultimately result in an increase in fiber transport by several percent, as shown in Fig.~\ref{fig:concentration_channel}(d). Interestingly, the dominant effect on flux modification is the concentration profile, despite the fact that the fibers are slightly slower than the fluid in the center of the channel. In particular, we observe that the enhancement only occurs for fibers whose lengths are within the log-layer of the wall (typically $\ell^+>25$). When $\ell^+\in[25,270]$, the increase seems proportional to $\log \ell^+$.

To conclude, we have observed an enhancement in the transport of inertialess flexible fibers in the case when they can be modeled as slender-bodies and interact with boundaries through a simple soft-boundary approach. However, many questions still remain. One such question is the need for more realistic models for interactions with boundaries to take into account the dissipation that can occur during collisions due to wall-normal lubrication forces, plastic deformations during contact and/or tangential friction. These interactions could affect the transfer of momentum toward the wall-normal direction and influence the migration of fibers into the bulk. This might result in an increase of their concentration near the boundary, as observed in the turbophoresis of inertial particles~\cite{brandt2022particle}. This actually brings out a second question as to whether similar behavior can be observed for fibers possessing inertia. Another issue that requires further exploration is the role of two-way coupling~\cite{Elgobashi2006} in pole-vaulting effects. As fibers deform, they can store some of the fluid kinetic energy, which they can restore later to the flow, potentially reducing the significance of pole-vaulting in the migration of fibers away from the wall. Additionally, in practical applications such as in the papermaking industry~\cite{lundell2011fluid}, fibers can display internal properties that fluctuate along their arc-length, such as thickness or flexibility. This could lead to mixed states upon collisions with boundaries, with partial hook and pole-vaulting effects, or even to their break-up into smaller fragments. All these aspects are kept for future work.

We benefited from discussions with Sof\`{\i}a Allende, who is warmly acknowledged. We are grateful to the OPAL infrastructure from Universit\'e C\^ote d'Azur for providing computational resources. This work received support from the UCA-JEDI Future Investments funded by the French government (grant no.\ ANR-15-IDEX-01) and from the Agence Nationale de la Recherche (grant no.\ ANR-21-CE30-0040-01).

\end{document}